\documentclass[preprint,aps,draft]{revtex4}

\usepackage{graphicx}
\usepackage{dcolumn}
\usepackage{bm}

\begin{document}

\setlength{\unitlength}{1mm}
\textwidth 15.0 true cm
\textheight 22.0 true cm
\headheight 0 cm
\headsep 0 cm
\topmargin 0.4 true in
\oddsidemargin 0.25 true in

\title{Consistency relation for single scalar inflation}

\author{Andrei Gruzinov}
 \affiliation{Center for Cosmology and Particle Physics, Department of Physics, New York University, NY 10003}

\date{June 4, 2004}

\begin{abstract}

Single scalar field inflation with a generic, non-quadratic in derivatives, field Lagrangian is considered. It is shown that non-Gaussianity of curvature perturbations is characterized by two dimensionless amplitudes. One of these amplitudes can be expressed in terms of the usual inflationary observables -- the scalar power, the tensor power, and the tensor index. This consistency relation provides an observational test for the single scalar inflation. 

\end{abstract}

\pacs{}

\maketitle

\section{Introduction}

It is hoped that the tentative status of inflationary theory will end when a consistency relation \cite{liddle} between the scalar power, the tensor power, and the tensor slope is tested observationally. For single scalar inflation with the field Lagrangian ${\cal L} ={1\over 2}(\partial \phi)^2-V(\phi )$, the consistency relation is $P_T/P_S=-8n_T$. The tensor power $P_T$, the scalar power $P_S$, and the tensor spectral index $n_T$ are inflationary observables.

If however a more general single field Lagrangian is considered, ${\cal L} ={\cal L}(\xi , \phi )$, with $\xi \equiv {1\over 2}(\partial \phi)^2$, the usual consistency relation should be replaced by \cite{mukh}
\begin{equation}\label{con1}
P_T/P_S=-8c_sn_T,
\end{equation}
where 
\begin{equation}
c_s^2\equiv {{\cal L}_\xi \over {\cal L}_\xi+2\xi {\cal L}_{\xi \xi}}
\end{equation}
is the speed of high-frequency scalar perturbations; ${\cal L}_\xi$ denotes partial derivative. The speed of sound $c_s$ cannot be measured independently. Thus this kind of inflation escapes the usual observational test. 

Recently Creminelli \cite{crem} noted that single scalar inflation with non-quadratic kinetic energy is characterized by a large (as compared to standard single scalar \cite{mald}) and therefore potentially observable \cite{sco}, non-Gaussianity. Creminelli \cite{crem} considered a specific non-quadratic Lagrangian, but it is straightforward to generalize his approach to a generic ${\cal L} ={\cal L}(\xi , \phi )$. We will show that the three-point correlator of curvature perturbations depends on two dimensionless amplitudes. One of these amplitudes can be expressed in terms of $c_s$, thus allowing an observational test of single scalar inflation. Of course, the very fact that the three-point function depends on just two dimensionless amplitudes provides a consistency check by itself. But measuring just an amplitude, with a given functional dependence on the triangle of the wavevectors should be an easier task \cite{zal}.

\section{Three-Point Correlator: Results}
We are interested in the curvature perturbation $\zeta$. By definition, on superhorizon scales, $e^{\zeta}$ is proportional to the scale factor on uniform energy hypersurface \cite{bond}. 

We write the two-point correlator of the curvature perturbation in the following form
\begin{equation}
\langle\zeta({\bf k})\zeta({\bf k'})\rangle =(2\pi )^3\delta ({\bf k}+{\bf k'})P_2(k).
\end{equation}
Here $k$ is the length of the spatial vector ${\bf k}$.
The three-point correlator can be written as
\begin{equation}
\langle\zeta({\bf k_1})\zeta({\bf k_2})\zeta({\bf k_3})\rangle =(2\pi )^3\delta ({\bf k_1}+{\bf k_2}+{\bf k_3})P_3(k_1,k_2,k_3).
\end{equation}

Define a dimensionless 3-point function
\begin{equation}
f(k_1,k_2,k_3)\equiv {P_3(k_1,k_2,k_3)\over P_2(k_1)P_2(k_2)+P_2(k_1)P_2(k_3)+P_2(k_2)P_2(k_3)}.
\end{equation}
The convenience of this definition is especially clear for post-inflationary non-Gaussianities. In this case $\zeta = g-(3/5)f_{nl}g^2$, where $g$ is a Gaussian random field and $f_{nl}$ is a constant, and we get a constant $f(k_1,k_2,k_3)=-(6/5)f_{nl}$. We expect that $f\sim 1$ might be observationally testable \cite{sco} .

We show in the Appendix that non-qudratic single scalar inflation gives the dimensionless 3-point function 
\begin{equation}\label{cor}
f(k_1,k_2,k_3)=f_1\left( {3\over 2}+{1\over 9}{p^2\over k^2}-{kp^2\over q^3}-{p^4\over kq^3}\right)~+~f_2{k_1^2k_2^2k_3^2\over k^3q^3}
\end{equation}
where
\begin{equation}
k\equiv{k_1+k_2+k_3\over 3},~~~~~~p^2\equiv{k_1^2+k_2^2+k_3^2\over 3},~~~~~~q^3\equiv{k_1^3+k_2^3+k_3^3\over 3},
\end{equation}
and the two amplitudes are
\begin{equation}\label{con2}
f_1={1-c_s^2\over c_s^2}
\end{equation}
\begin{equation}
f_2={2\over 27}{(1-c_s^2)^2\over c_s^2}-{4\over 81}(1-c_s^2){\xi {\cal L}_{\xi \xi \xi}\over {\cal L}_{\xi \xi }}.
\end{equation}
Equations (\ref{con1}) and (\ref{con2}) form a consistency relation for single scalar inflation. 

As in \cite{crem}, the above expressions are valid only in the leading order in the slow roll parameters. For degenerate triangles with $k_1\approx k_2\gg k_3$, the 3-point function $f(k_1,k_2,k_3)$ vanishes as $k_3^2$. It is also straightforward to check that our formulas reproduce the 3-point function of \cite{crem} in the leading order in $1-c_s^2$ and $\xi {\cal L}_{\xi \xi \xi}/{\cal L}_{\xi \xi }$.

\section{Discussion}

The 3-point function for single scalar inflation with non-quadratic Lagrangian is given by two dimensionless amplitudes. One of these amplitudes can be expressed in terms of the scalar power, the tensor power, and the tensor spectral index. This provides a consistency relation.

\begin{acknowledgments}
This work was supported by the David and Lucile Packard Foundation.
\end{acknowledgments}

\begin{appendix}

\section{2-point correlator}
We consider field Lagrangian ${\cal L} ={\cal L}(\xi , \phi )$, with $\xi \equiv {1\over 2}(\partial \phi)^2$. In the leading order in slow roll parameters, one can calculate the correlators of $\phi$ in de-Sitter space, and then translate these into the correlators of $\zeta$ by a linear relation \cite{mald,crem}
\begin{equation}\label{pz}
\zeta=-{H\over \dot{\phi _0}}\phi ={H\over \sqrt{2\xi }}\phi,
\end{equation}
where $\dot{\phi _0}=-\sqrt{2\xi }$ is the unperturbed value, and $\phi$ is the perturbation. 

The second order Lagrangian for $\phi$ is
\begin{equation}
\delta _2{\cal L}=a^3\left( {\cal L}_{\xi }\delta _2\xi +{1\over 2}{\cal L}_{\xi \xi }(\delta _1\xi )^2\right).
\end{equation}
Here 
\begin{equation}
\delta _1\xi=\dot{\phi _0}\dot{\phi },~~~~~~~\delta _2\xi={1\over 2}\dot{\phi }^2-{1\over 2a^2}(\partial _i\phi )^2.
\end{equation}
and $a=e^{Ht}$ is the scale factor. Introducing the speed of sound 
\begin{equation}
c_s^2\equiv {{\cal L}_\xi \over {\cal L}_\xi+2\xi {\cal L}_{\xi \xi}},
\end{equation}
we can write the second order Lagrangian as
\begin{equation}
\delta _2{\cal L}={{\cal L}_\xi \over c_s^2}{a^3\over 2}\left( \dot{\phi }^2-{c_s^2\over a^2}(\partial _i\phi )^2\right) .
\end{equation}
This gives 
\begin{equation}
\langle\phi({\bf k})\phi({\bf k'})\rangle =(2\pi )^3\delta ({\bf k}+{\bf k'}){H^2\over 2k^3}{1\over c_s{\cal L}_\xi}
\end{equation}
and 
\begin{equation}
\langle\zeta({\bf k})\zeta({\bf k'})\rangle =(2\pi )^3\delta ({\bf k}+{\bf k'}){H^2\over 2k^3}{H^2\over 2c_s\xi {\cal L}_\xi} ,
\end{equation}
which is the result of \cite{mukh}.

\section{3-point correlator}

The cubic Lagrangian is 
\begin{equation}
\delta _3{\cal L}=a^3\left( {\cal L}_{\xi \xi}\delta _1\xi \delta _2\xi +{1\over 6}{\cal L}_{\xi \xi \xi}(\delta _1\xi )^3\right) 
\end{equation}
\begin{equation}
=-\sqrt{2\xi }a^3\left( ({1\over 2}{\cal L}_{\xi \xi}+{1\over 3}\xi {\cal L}_{\xi \xi \xi})\dot{\phi }^3-{1\over 2a^2}{\cal L}_{\xi \xi }\dot{\phi } (\partial _i\phi )^2  \right) 
\end{equation}

As in \cite{mald,crem}, the $\phi$ 3-point correlator is calculated from
\begin{equation}
\langle\phi ^3\rangle =-i\int dt \langle [ L_3(t),\phi^3] \rangle,
\end{equation}
using the 2-point correlators
\begin{equation}
\langle\phi({\bf k},\eta )\phi({\bf k'},\eta =0)\rangle =(2\pi )^3\delta ({\bf k}+{\bf k'}){H^2\over 2k^3}{1\over c_s{\cal L}_\xi}e^{-ic_sk\eta }(1+ic_sk\eta ).
\end{equation}
Here $\eta =-e^{-Ht}/H$ is conformal time, and taking the limiting value $\eta =0$ in the correlator corresponds to calculating the 3-point function after horizon crossing. Translation into the $\zeta$ correlator is done using (\ref{pz}). The result can be presented in the form (\ref{cor}).

\end{appendix}

\end{document}